\overfullrule=0pt
\input harvmac
\def\a{{\alpha}}
\def\ad{{\dot a}}
\def\Ad{{\dot A}}
\def\ah{{\widehat \a}}

\def\l{{\lambda}}

\def\b{{\beta}}
\def\bt{{\widetilde\beta}}

\def\g{{\gamma}}
\def\G{{\Gamma}}
\def\gt{{\widetilde\gamma}}

\def\d{{\delta}}
\def\e{{\epsilon}}
\def\s{{\sigma}}
\def\k{{\kappa}}

\def\half{{1\over 2}}
\def\p{{\partial}}
\def\pb{{\overline\partial}}
\def\pt{{\partial_\tau}}
\def\t{{\theta}}

\def\bar{\overline}

\Title{\vbox{\hbox{IFT-P.019/2007 }}}
{\vbox{
\centerline{\bf Explaining the Pure Spinor Formalism for the Superstring
}}}
\bigskip\centerline{Nathan Berkovits\foot{e-mail: nberkovi@ift.unesp.br}}
\bigskip
\centerline{\it Instituto de F\'\i sica Te\'orica, State University
of S\~ao Paulo}
\centerline{\it Rua Pamplona 145, 01405-900, S\~ao Paulo, SP, Brasil}

\vskip .3in
After adding a pair of non-minimal fields and performing a similarity transformation, 
the BRST
operator in the pure spinor formalism
is expressed as a conventional-looking BRST operator involving
the Virasoro constraint and $(b,c)$ ghosts, together with 12 fermionic 
constraints. This BRST operator can be obtained by gauge-fixing the
Green-Schwarz superstring where the 8 first-class and 8 second-class
Green-Schwarz constraints are combined into 12 first-class constraints.
Alternatively, the pure spinor BRST operator can be obtained from the 
RNS formalism by twisting the ten spin-half RNS fermions into five spin-one and 
five spin-zero fermions, and using the $SO(10)/U(5)$ pure spinor variables 
to parameterize the different ways of twisting. $GSO(-)$ vertex operators in 
the pure spinor formalism are constructed using spin fields and 
picture-changing operators in a manner analogous to Ramond vertex operators 
in the RNS formalism.

\vskip .3in

\Date {December 2007}

\newsec{Introduction}

The pure spinor formalism \ref\psf{N. Berkovits, {\it
Super-Poincar\'e covariant quantization of the superstring},
JHEP 0004 (2000) 018, hep-th/0001035.} is a super-Poincar\'e covariant
description of the superstring
which significantly simplifies multiloop amplitude computations and
which allows quantization in Ramond-Ramond backgrounds.
However, because of the non-conventional form of the BRST operator in the pure
spinor formalism, the relation of this formalism to the Green-Schwarz (GS) and
Ramond-Neveu-Schwarz (RNS) formalisms for the superstring was mysterious. Furthermore,
it was not known how to describe the $GSO(-)$ sector of the superstring using the
pure spinor formalism. 

In this paper, these mysterious features of the formalism will be explained by
adding a pair of non-minimal fields and performing a similarity transformation such
that the pure spinor BRST operator is expressed as a conventional-looking BRST
operator. This conventional-looking BRST operator involves the Virasoro constraint 
and twelve fermionic constraints, where eleven
of these fermionic constraints are associated to the eleven independent
components of the original bosonic pure spinor ghost. 
The twelfth fermionic constraint and the Virasoro constraint
are associated to the new pair of non-minimal fields, $(\bt,\gt)$ and $(b,c)$, which
have opposite statistics and carry conformal weight $(2,-1)$.
Although this conventional form of the BRST operator is not manifestly Lorentz
invariant, it will be useful for constructing $GSO(-)$ vertex operators and for
relating the pure spinor formalism to the GS and RNS formalisms.

The new non-minimal fields, $(\bt,\gt)$ and $(b,c)$,
decouple from vertex operators and scattering amplitudes
involving $GSO(+)$ states, however, they play a crucial role in defining
vertex operators and scattering amplitudes involving $GSO(-)$ states. Just as
Ramond vertex operators in the RNS formalism \ref\fms{D. Friedan, E. Martinec
and S. Shenker, {\it Conformal invariance, supersymmetry and string theory},
Nucl. Phys. B271 (1986) 93.}
depend non-trivially on the $(\b,\g)$ ghosts,
$GSO(-)$ vertex operators in the pure spinor formalism will depend non-trivially on
the $(\bt,\gt)$ ghosts. And just as scattering amplitudes involving Ramond states in
the RNS formalism
require picture-changing operators to cancel the picture of the Ramond vertex operators,
scattering amplitudes involving $GSO(-)$ states in the pure spinor formalism will require
picture-changing operators to cancel the picture of the $GSO(-)$ vertex operators.

Note that the new non-minimal fields $(\bt,\gt,b,c)$
are unrelated to the non-minimal fields $(\bar\l_\a, \bar w^\a, r_\a,s^\a)$
which were introduced in the ``Dolbeault'' description of the pure spinor formalism
\ref\topo{N. Berkovits, {\it Pure spinor formalism as an $N=2$ topological string},
JHEP 0510 (2005) 089, hep-th/0509120.}\ref\nikj{N. Nekrasov,
{\it Lectures at the 23rd Jerusalem Winter School in Theoretical Physics},
January 2006,
http://www.as.huji.ac.il/schools/phys23/media2.shtml.}
\ref\nikme{N. Berkovits and N. Nekrasov,
{\it  Multiloop superstring amplitudes from non-minimal pure spinor formalism},
JHEP 0612 (2006) 029, hep-th/0609012.}.
In this paper, the Dolbeault description will not be discussed although it would 
be interesting to consider including both 
$(\bt,\gt,b,c)$
and $(\bar\l_\a, \bar w^\a, r_\a,s^\a)$ non-minimal fields in the pure spinor formalism. Such
a Dolbeault description might be useful for writing the conventional-looking BRST operator
in a manifestly Lorentz-invariant form.

After expressing the pure spinor BRST operator as 
a conventional-looking BRST operator with a Virasoro constraint and twelve
fermionic constraints, it is relatively straightforward to relate the pure 
spinor formalism with the GS and RNS formalisms for the superstring.
In the GS formalism, the fermionic constraint $d_\a=0$ contains 8 first-class
components and 8 second-class components. 
After breaking manifest Lorentz invariance down to $SO(8)$ and then to $U(4)$, 
the 8 second-class constraints can be converted into 4 first-class constraints.
The resulting BRST operator has 12 fermionic constraints and is related by
a field redefinition to the pure spinor BRST operator. Interestingly,
this field redefinition allows the manifest $U(4)$ invariance to 
be enlarged to $U(5)$.\foot{Throughout this paper, we shall Wick-rotate the
spacetime metric from $SO(9,1)$ to $SO(10)$. All results can be Wick-rotated 
back to Minkowski space, however, the group structure of the 25 $U(5)$ generators
is more complicated in Minkowski space where it splits into 16 $U(4)$ generators
and 9 light-like boosts.}

To relate the RNS formalism with the pure spinor formalism, one first twists the ten
spin-half RNS fermions $\psi^m$ into five spin-zero fermions $\t^a$ and five
spin-one fermions $p_a$ for $a=1$ to 5.\foot
{A similar twisting procedure 
was used in several earlier papers by Baulieu and collaborators to
relate the RNS string to a topological string \ref\baulieu
{L. Baulieu, {\it Transmutation of pure 2-D supergravity into topological 2-D
gravity and other conformal theories}, Phys. Lett. B288 (1992) 59,
hep-th/9206019\semi
L. Baulieu, M.B. Green and E. Rabinovici, {\it
A unifying topological action for heterotic and type II
superstring theories}, Phys. Lett. B386 (1996) 91, hep-th/9606080\semi
L. Baulieu and N. Ohta, {\it Worldsheets with extended supersymmetry},
Phys. Lett. B391 (1997) 295, hep-th/9609207\semi
L. Baulieu, M.B. Green and E. Rabinovici, {\it
Superstrings from theories with $N>1$ worldsheet
supersymmetry}, Nucl. Phys. B498 (1997) 119,
hep-th/9611136.}. I 
thank Nikita Nekrasov for informing me of these papers.}
This twisting breaks $SO(10)$ Lorentz
invariance to $U(5)$, and one can parameterize the different choices of twisting by
introducing $SO(10)/U(5)$ bosonic pure spinor variables. One then imposes the
constraints that physical states are independent of the 11 pure spinor variables,
and the fermionic ghosts for these constraints are the remaining 11 components of
$\t^\a$ and $p_\a$. After adding these 11 constraints
to the N=1 super-Virasoro constraints, the RNS BRST operator is mapped into the conventional
form of the pure spinor BRST operator where the spin $-1$ non-minimal field $\gt$
is related to the spin $-\half$ RNS ghost $\g$ as $\gt= (\g)^2$.

It is interesting to note that a similar procedure of twisting fermions has been used
to embed the $N=0$ bosonic string into an $N=1$ string 
\ref\unique{N. Berkovits and C. Vafa, {\it On the uniqueness of string theory},
Mod. Phys. Lett. A9 (1994) 653, hep-th/9310170.}. In the $N=0\to N=1$ embedding,
the $(b,c)$ ghosts are twisted from $(2,-1)$ conformal weight to $({3\over 2},-\half)$
conformal weight, and the $N=1$ stress tensor is defined as $G=b + j_{BRST}$ where
$\int dz ~j_{BRST}$ is the BRST charge of the bosonic string.
In fact, the inverse map of this embedding which takes an $N=1$ string into an
$N=0$ string is closely related to the map from the RNS formalism to the pure
spinor formalism. This is not surprising since the pure spinor formalism can be
interpreted as an $N=2$ topological string \topo
\ref\sken{J. Hoogeveen and
K. Skenderis, {\it BRST quantization of the pure spinor
superstring}, arXiv:0710.2598 [hep-th].},
which is a natural generalization of
$N=0$ bosonic strings.

Note that this $N=1\to N=0$ inverse map from the RNS to the pure spinor formalism
is different from the $N=1\to N=2$ embedding that has been used to map the RNS
formalism into the hybrid formalism for the superstring \ref\newdes{N. Berkovits,
{\it A new description of the superstring}, hep-th/9604123.}. 
For example, the $N=1\to N=2$
embedding maps the RNS string into a critical $\hat c=2$ $N=2$ string as opposed
to a $\hat c=3$ $N=2$ topological
string. 
Nevertheless, a certain version of the $N=1\to N=2$ embedding
will be shown at the end of this paper
to closely resemble the $N=1\to N=0$ embedding. This version might eventually
be useful 
for relating the pure spinor and hybrid formalisms for the superstring.

In section 2 of this paper, the pure spinor formalism is briefly reviewed and a
pair of non-minimal fields, $(b,c)$ and $(\bt,\gt)$, are introduced.
After performing a similarity transformation, the pure spinor BRST operator is
expressed as a conventional-looking BRST operator with 12 fermionic constraints.

In section 3, $GSO(-)$ vertex operators are constructed with the help
of the non-minimal fields. These $GSO(-)$ vertex operators carry nonzero picture
and, after defining picture-changing operators, it is shown
how to compute scattering amplitudes using these vertex operators.

In section 4, the conventional-looking form of the pure spinor BRST operator is
obtained from gauge-fixing the GS superstring. In performing this gauge-fixing,
the 8 first-class and 8 second-class GS constraints are combined into 12 first-class
constraints in a manifestly $U(5)$-invariant manner.

In section 5, the RNS BRST operator is mapped to the pure spinor BRST operator
by twisting the ten spin-half RNS fermions using an $SO(10)/U(5)$ pure spinor
variable to parameterize the different twistings. For states in the Neveu-Schwarz
$GSO(+)$ sector, it is shown how to map the RNS and pure spinor vertex operators into
each other.

In section 6, the map from the RNS formalism to the pure spinor formalism is
interpreted as an inverse map of the $N=0\to N=1$ embedding of the bosonic string.
This inverse map may be useful for constructing generalizations of the pure spinor
formalism.

And in section 7, the approach of this paper will be compared with other
approaches to ``explaining'' the pure spinor formalism. One approach which is discussed
in detail uses an $N=1\to N=2$ embedding to map the RNS string into variables which
closely resemble those of the $N=1\to N=0$ embedding.

\newsec{Conventional-Looking Pure Spinor BRST Operator}

In this section, the pure spinor formalism will be briefly reviewed
and the BRST operator $Q=\int dz \l^\a d_\a$
will be related to a conventional-looking
BRST operator involving the usual $(b,c)$ ghosts and Virasoro constraint,
together with 12 fermionic constraints.

\subsec{Brief review of pure spinor formalism}

The pure spinor formalism \psf\
in a flat background is described by the free worldsheet action
\eqn\actionp{S=\int d^2 z [ \half \p x^m \pb x_m + p_\a \pb \t^\a + \widehat p_\ah
\p \widehat \t^\ah + w_\a \pb \l^\a + \widehat w_\ah \p \widehat \l^\ah]}
where $(x^m,\t^\a,\widehat\t^\ah,p_\a,\widehat p_\ah)$ are the Green-Schwarz-Siegel
matter variables for $m=0$ to 9 and $(\a,\ah)=1$ to 16,
$(\l^\a,w_\a)$ and $(\widehat\l^\ah,\widehat w_\ah)$ are left and right-moving
bosonic ghost variables satisfying the pure spinor constraint
\eqn\puredeff{\l\G^m\l= \widehat\l\G^m\widehat\l=0,}
and $\G^m_{\a\b}$ and $(\G^m)^{\a\b}$ are $16\times 16$ symmetric matrices satisfying 
$\G^{(m}_{\a\b}(\G^{n)})^{\b\g} =2 \d_\a^\g \eta^{mn}$.
The hatted spinor variables have the opposite/same chirality as the unhatted variables
for the Type IIA/IIB superstring, and throughout this paper, the hatted variables
will be ignored.

Physical states are defined as states in the cohomology of the BRST operator
\eqn\brstone{Q = \int dz~ \l^\a d_\a}
where
\eqn\dddef{d_\a = p_\a -\half (\G^m\t)_\a \p x_m -{1\over 8} (\t\G^m \p\t)(\G_m\t)_\a}
is the Green-Schwarz constraint. Since
$d_\a$ satisfies the OPE 
\ref\siegeltwo{W. Siegel, {\it Classical superstring mechanics},
Nucl. Phys. B263 (1985) 93.}
\eqn\dope{d_\a(y) d_\b(z)\to -(y-z)^{-1} \G^m_{\a\b}\Pi_m}
where $\Pi_m = \p x_m +\half\t\G_m\p \t $ is the supersymmetric momentum, 
$Q$ is nilpotent using the constraint of \puredeff.

For massless super-Yang-Mills states, the 
unintegrated and integrated vertex operators are
\eqn\unint{V = \l^\a A_\a(x,\t),}
$$
\int dz ~U = \int dz [
\p\t^\a A_\a(x,\t) + \Pi^m A_m(x,\t) + d_\a W^\a(x,\t) 
+ N_{mn} F^{mn}(x,\t)]$$
where 
$N^{mn}=\half (w\G^{mn}\l)$ is the Lorentz current for the pure spinor variables,
$(A_\a,A_m)$ are gauge superfields and $(W^\a,F^{mn})$ are 
superfield-strengths for super-Yang-Mills.
When the super-Yang-Mills superfields are onshell, $QV=0$ and $QU = \p V$.

Tree-level $N$-point scattering amplitudes are computed by the correlation function
\eqn\treea{{\cal A} = \langle\langle V_1 V_2 V_3 ~\int dz_4 U_4 ... 
\int dz_N U_N \rangle\rangle}
using the measure factor
\eqn\measure{\langle\langle (\l\G^m\t)(\l\G^n\t)(\l\G^p\t)(\t\G_{mnp}\t)\rangle\rangle =1.}
Although this measure factor looks unusual, it can be derived from functional integration
over the worldsheet fields after performing a BRST-invariant regularization 
\topo.

The correlation function of 
\treea\ is easily computed using the free-field OPE's coming from the worldsheet action of
\actionp\ together with the OPE's 
\eqn\open{N^{mn}(y)\l^\a(z) \to \half (y-z)^{-1} (\G^{mn}\l)^\a,}
$$N^{mn}(y) N^{pq}(z)\to (y-z)^{-1}(\eta^{p[n} N^{m]q}
-\eta^{q[n} N^{m]p}) - 3 (y-z)^{-2} \eta^{m[q}\eta^{p]n}.$$
The manifestly covariant
OPE's of \open\ can be derived by solving the pure spinor constraint $\l\G^m\l=0$
in a $U(5)$-invariant manner. Under $SU(5)\times U(1)$, an $SO(10)$ spinor decomposes
as $\l^\a \to (\l^+, \l_{ab},\l^a)$ where $a=1$ to 5, $\l_{ab}=-\l_{ba}$, and
$(\l^+,\l_{ab},\l^a)$ carries $U(1)$ charge $({5\over 2}, \half, -{3\over 2})$.
If $\l^+$ is assumed to be nonzero, $\l\G^m\l=0$ implies that 
\eqn\solll{\l^a = -{1\over 8} (\l^+)^{-1} \e^{abcde}\l_{bc}\l_{de}}
so that $\l^\a$ has eleven independent components parameterized by $\l^+$ and $\l_{ab}$.

In terms of $(\l^+,\l_{ab})$ and their conjugate momenta $(w_+,w^{ab})$, 
the pure spinor contribution to the stress tensor and Lorentz currents is
\ref\relatr{N. Berkovits, {\it
Relating the RNS and pure spinor formalisms for the superstring}, JHEP 0108 (2001) 026,
hep-th/0104247.}
\eqn\tpure{T_{pure} = \half w^{ab}\p\l_{ab} + w_+\p\l^+ +{3\over 2}
\p^2 (\log ~\l^+),}
\eqn\lorentz{ N_{U(1)} = {1\over{\sqrt 5}}({1\over 4}
\l_{ab} w^{ab} + {5\over 2} \l^+ w_+ - {5\over 4} \p (\log ~\l^+)),}
$$N^{ab} = \l^+ w^{ab},\quad N_a^b = \l_{ac} w^{bc} -{1\over 5}\d_a^b
\l_{cd} w^{cd},$$
$$
N_{ab} = (\l^+)^{-1} ( 2 \p\l_{ab}  -{5\over 2}\l_{ab}\p(\log~\l^+)
+ \l_{ac}\l_{bd} w^{cd} -\half \l_{ab}\l_{cd} w^{cd}) - w_+ \l_{ab},$$
where the $SO(10)$ Lorentz currents $N^{mn}$
have been decomposed into $(N_{U(1)}, N_a^b, N^{ab}, N_{ab})$
which transform as $(1,24,10,\bar{10})$ representations of $SU(5)$.
Note that the ``improvement'' term ${3\over 2}\p^2 (\log \l^+)$ is necessary in $T_{pure}$
so that $N_{mn}$ are primary fields with respect to $T_{pure}$.
It is also convenient to 
define the ghost-current 
\eqn\ghost{J = w_+\l^+ +\half w^{ab}\l_{ab} + {7\over 2}\p(\log ~\l^+)}
which has no poles with $N^{mn}$ and which satisfies $J(y)\l^\a(z)\to (y-z)^{-1}\l^\a$.

Although there is no fundamental $b$ ghost in the pure spinor formalism,
one can construct a composite operator $G^\a$ satisfying $\{Q, G^\a\} = \l^\a T$
where 
\eqn\tstress{T= -\half \p x^m \p x_m - p_\a \p\t^\a + T_{pure}}
is the stress tensor with zero central charge.\foot{It is interesting
to point out that in a curved target-space background,
$G^\a$ will in general not be holomorphic. Nevertheless, one can argue
that $\bar\p G^\a$ is BRST-trivial, which appears to be sufficient
for computing scattering amplitudes where $G^\a$ plays the role of
the $b$ ghost.}
This composite operator will play an important role
in this paper and is defined as \relatr
\eqn\defgg{G^\a =\half \Pi^m (\G_m d)^\a 
-{1\over 4} N^{mn} (\G_{mn}\p\t)^\a -{1\over 4} 
J \p\t^\a -{1\over 4}\p^2\t^\a}
where $N_{mn}$ and $J$ are defined in \lorentz\ and \ghost. 

\subsec{Non-minimal fields and similarity transformation}

The first step to constructing a conventional-looking
BRST operator from $Q=\int dz \l^\a d_\a$ is to
add the term 
$\int dz \gt b$ to the pure spinor BRST operator so that
\eqn\firsts{Q = \int dz (\l^\a d_\a + \gt b)}
where $(\bt,\gt)$ are bosonic and $(b,c)$ are fermionic non-minimal
fields with the worldsheet action
$\int d^2 z (\bt \pb \gt + b \pb c)$. These non-minimal fields do not contribute
to the cohomology because of the topological term $\int dz \gt b$ in $Q$.

The second step is to perform the similarity transformation
$Q' = e^R Q e^{-R}$ where
\eqn\defR{R = \oint dz [ -{c\over {\l^+}} G^+ + c \p c \bt]}
and $G^+$ is
the component of $G^\a$ in \defgg\ with $5\over 2$ $U(1)$ charge. Using
$\{Q, G^+\}=\l^+ T$, it
is easy to verify that after performing the similarity transformation,
\eqn\newq{Q' = 
e^R Q e^{-R} =\int dz [ c \widetilde T - {\gt\over{\l^+}} G^+ + \l^\a d_\a +
\gt b + b c \p c ]}
where 
\eqn\newT{ \widetilde T = -\half \p x^m \p x_m - p_\a \p\t^\a + T_{pure}
+ \bt\p\gt + \p(\bt\gt)}
is a stress tensor with central charge $c=26$.

Although $Q'$ is not invariant under Lorentz transformations generated by $M_{ab}$
which transform $\l^+$ and $G^+$ into $\l_{ab}$ and $G_{ab}$, one can use the
relation \topo
\eqn\relhh{\l^{[\a} G^{\b]} = [ \int dz ~\l^\g d_\g,~H^{\a\b}~] }
where
\eqn\defhh{H^{\a\b} = {1\over{192}}\G_{mnp}^{\a\b} (d \G^{mnp} d + 24 N^{mn}\Pi^p)}
to show that 
\eqn\qlorentz{[~Q', ~M_{ab} + \int dz {\gt \over{(\l^+)^2}} H_{ab}^+ ~]=0}
where $H^+_{ab}$ is the component of $H^{\a\b}$
with $\a = +$ and $\b = ab$. Furthermore, one can verify that
the Lorentz algebra generated by $M'_{ab}\equiv M_{ab}+
\int dz {\gt \over{(\l^+)^2}} H_{ab}^+ $ with the other Lorentz generators
closes up to a BRST-trivial operator.
So under Lorentz transformations generated by $M'_{mn}=[M_{U(1)}, M^{ab}, M^a_b, M_{ab}
+\int dz {\gt\over{(\l^+)^2}} H_{ab}]$, $Q'$-closed states transform covariantly up to a BRST-trivial
transformation. Note that one could have defined the Lorentz generators
as $M'_{mn} = e^R M_{mn} e^{-R}$ where $R$ is defined in \defR, but such a definition
would not preserve the property that all poles when $\l^+\to 0$ have residues which 
are proportional to $\gt$. As will be discussed later, this property
is useful since terms proportional to $\gt$ will decouple from scattering amplitudes.

Finally, it will be convenient to define 
\eqn\defgt{\gt_+ =- {\gt\over{\l^+}},}
so that 
\eqn\nnq{Q' = \int dz [ c \widetilde T + \gt_+ G^+ + \l^\a d_\a -
\l^+ \gt_+ b + b c \p c ].}
If $(\gt_+, \l^+, \l_{ab})$ are interpreted as 12 independent
bosonic ghosts, $Q'$ resembles a standard BRST operator constructed
from 12 fermionic constraints and the Virasoro constraint.

Since $(\gt_+,\bt^+)$ are not Lorentz scalars, they will appear in
the Lorentz generators. In terms of $(\gt_+,\bt^+)$, $(w_+,\l^+)$
and $(w^{ab},\l_{ab})$, the $SO(10)$ Lorentz currents of \lorentz\ are
\eqn\lorentznew{ N_{U(1)} = {1\over{\sqrt 5}}({1\over 4}
\l_{ab} w^{ab} + {5\over 2} \l^+ w_+ - {5\over 2} \gt_+ \bt^+),}
$$N^{ab} = \l^+ w^{ab},\quad N_a^b = \l_{ac} w^{bc} -{1\over 5}\d_a^b
\l_{cd} w^{cd},$$
$$
N_{ab} = (\l^+)^{-1} ( 2 \p\l_{ab} + \l_{ab} \gt_+\bt^+ -4\l_{ab}\p(\log~\l^+)
+ \l_{ac}\l_{bd} w^{cd} -\half \l_{ab}\l_{cd} w^{cd}) - w_+ \l_{ab}.$$

The contribution of these bosonic ghosts to the stress tensor is
\eqn\stressnew{
\widetilde T_{pure} = w_+\p\l^+ + \half w^{ab} \p\l_{ab} + \bt^+ \p\gt_+ + \p(\bt^+\gt_+),}
which can be verified to have no triple poles with $N^{mn}$.
And the ghost current of \ghost\ is
\eqn\ghostnew{ J = w_+\l^+ + \half w^{ab}\p\l_{ab} - \bt^+\gt_+ + 4 \p(\log~\l^+).}
Remarkably, after including the $(b,c)$ and $(\bt^+,\gt_+)$ non-minimal
fields, $\widetilde T_{pure}$ no longer requires improvement terms involving
$\p^2 (\log ~\l^+)$. This may resolve some of the puzzles discussed in \ref\nikl
{N. Nekrasov, {\it Lectures on curved beta-gamma systems, pure spinors, and anomalies},
hep-th/0511008.} 
which are related to possible anomalies in the formalism. Furthermore, 
as will be
shown in the following section, the introduction of these non-minimal
fields appears to be necessary for the construction of $GSO(-)$ vertex operators
in the pure spinor formalism.

\newsec{ $GSO(-)$ States in the Pure Spinor Formalism}

In this section, it will be shown how to construct vertex operators
for $GSO(-)$ states and, after defining picture-changing operators,
it will be shown how to compute scattering amplitudes involving these states.

\subsec{$GSO(+)$ vertex operators}

Before constructing $GSO(-)$ vertex operators, it will be useful
to explain how to construct $GSO(+)$ vertex operators using the
new BRST operator $Q'$ of \nnq.
For $GSO(+)$ states, one method to construct vertex
operators $V'$ which are BRST-invariant with respect to $Q'$ 
is to simply define $V' = e^R V e^{-R}$ where $R$ is defined in
\defR\ and $V$ is the original pure spinor vertex operator which
is BRST-invariant with respect to $Q=\int dz \l^\a d_\a$.

However, a more useful definition is
\eqn\vpr{V' = c U + V + \gt_+ (G_0^+ U) + {{c\gt_+}\over{\l^+}}
(G_{-1}^+ G_0^+ U)}
where $V$ and $\int dz U$ are the original pure spinor unintegrated
and integrated vertex operators satisfying $QV=0$ and $QU = \p V$,
$G^+_{n}$ signifes the pole of order $(n+2)$ with $G^+$,
and $V$ has been gauge-fixed to satisfy $G_n^+ V=0$ for $n\geq 0$.
For example, for the massless super-Yang-Mills vertex operator
$V$ and $U$ of \unint,
the gauge-fixing condition $G_0^+ V =0$
implies that $\p_m (\g^m D)^+ A_\a =0$, which implies that
$\p^m \p_m A_\a = \p^m A_m =0$.

Note that $V'$ of \vpr\ is related to $e^R V e^{-R}$
by the BRST-trivial transformation
\eqn\brs{V' = e^R V e^{-R} - Q' ({c\over \l^+} G_0^+ U)}
where the relation
\eqn\rela{Q G^+_0 U = -G^+_0 QU + \l^+ T_0 U = - G_0^+ \p V + \l^+ U
= - \p (G^+_0 V) + G^+_{-1} V + \l^+ U}
has been used.
Although both \vpr\ and $e^R V e^{-R}$ have poles when $\l^+\to 0$, the
vertex operator of \vpr\ has the advantage that the residues of these poles
are proportional to $\gt_+$. Since the vertex operators are independent of $\bt^+$,
any term proportional to $\gt_+$ will generically decouple from scattering amplitudes.

\subsec{ $GSO(-)$ vertex operators}

For $GSO(-)$ states, it does not appear to be possible to construct
vertex operators in the original pure spinor formalism without
the non-minimal $(\bt^+,\gt_+)$ fields \ref\partha
{P. Mukhopadhyay, {\it DDF construction and D-brane boundary
states in pure spinor formalism}, JHEP 0605 (2006) 055, hep-th/0512161.}
\ref\schiappa{R. Schiappa and N. Wyllard, {\it D-brane boundary
states in the pure spinor superstring}, JHEP 0507 (2005) 070,
hep-th/0503123.}. 
The reason
is that, just as Ramond vertex operators in the RNS formalism depend
non-trivially on the $(\b,\g)$ ghosts, the $GSO(-)$
vertex operators in the pure spinor formalism will depend non-trivially
on the $(\bt^+,\gt_+)$ ghosts.

For example, the tachyon vertex operator in the pure spinor formalism will
be 
\eqn\tach{V' = c \exp [-\half(3\widetilde\phi + \phi^+ + 
\sum_{[ab]=1}^{10}\phi_{ab} - i\sum_{\a=1}^{16} \s_\a)] e^{ik_m x^m}}
where the $(\bt^+,\gt_+)$, $(w_+,\l^+)$ and $(w^{ab},\l_{ab})$ bosonic
ghosts
have been fermionized as 
\eqn\ferm{\bt^+ = e^{-\widetilde\phi} \p\widetilde\xi^+,\quad 
\gt_+ = \widetilde\eta_+ e^{\widetilde\phi},}
$$w_+ = e^{-\phi^+} \p\xi_+,\quad 
\l^+ = \eta^+ e^{\phi^+},$$
$$w^{ab} = e^{-\phi_{ab}} \p\xi^{ab},\quad 
\l_{ab} = \eta_{ab} e^{\phi_{ab}},$$
and the $(\t^\a,p_\a)$ fields have been bosonized as
$$p_\a = e^{-i\s^\a}, \quad \t^\a = e^{i\s^\a}.$$
Since $e^{n\widetilde\phi}$ carries conformal weight $\half(-n^2-3n)$,
$e^{n\phi^+}$ and $e^{n\phi^{ab}}$ carry conformal weight $\half(-n^2-n)$,
and $e^{in\sigma_\a}$ carries conformal weight $\half(n^2 -n)$, one
finds that $V'$ of \tach\ carries zero conformal weight when
$e^{ik_m x^m}$ has $\half$ conformal weight as expected for the tachyon.
Furthermore, it is not difficult to show that $Q' V'=0$.

Although only $U(5)$ invariance is manifest, one can easily verify that \tach\
is a scalar under Lorentz transformations generated by \lorentznew. It is interesting
to note that bosonized Ramond vertex operators in the RNS formalism also manifestly
preserve only a $U(5)$ subgroup of the Lorentz group.

Other $GSO(-)$ vertex operators can be constructed by taking 
OPE's of the tachyon vertex operator of \tach\ with the $GSO(+)$
vertex operators of \vpr.
Just as $(\psi^m,\b,\g)$ have square-root cuts with Ramond vertex 
operators in the RNS formalism, $(\t^\a,p_\a,\l^\a,w_\a,\bt^+,\gt_+)$
have square-root cutes with $GSO(-)$ vertex operators in the pure spinor
formalism. 
To be convinced that this construction of $GSO(-)$ vertex operators
is correct, it
will now be shown how to compute tree amplitudes using these $GSO(-)$
vertex operators.

\subsec{Picture-changing operators}

Because of the screening charges related to
the conformal weights of the worldsheet fields,
the natural measure factor for tree amplitudes is
\eqn\nat{\langle c \p c \p^2 c~ (\t)^{16}~
\exp [-3\widetilde\phi -\phi^+ -\sum_{ab=1}^{10}\phi^{ab}] \rangle =1.}
If one defines picture such that $\xi$ and $e^\phi$ carry picture $+1$ and
$\eta$ carries picture $-1$,
the measure factor of \nat\ carries 
picture $(-3,-1,-1)$ with respect to the $(\gt_+,\l^+,\l_{ab})$ ghosts,
the $GSO(+)$ vertex operators of \vpr\ carry picture $(0,0,0)$, and
the $GSO(-)$ vertex operators of \tach\ carry picture $(-{3\over 2},-\half,-\half)$. 

To relate the measure factor of \nat\ to the usual pure spinor
measure factor
\eqn\purem{\langle\langle
(\l\g^m\t)(\l\g^n\t)(\l\g^p\t)(\t\g_{mnp}\t)\rangle\rangle =1}
which carries zero picture, one needs to introduce BRST-invariant
picture-raising
operators. As in the RNS formalism, the picture-raising operators
are naturally defined by anticommuting the BRST operator with the $\xi$
variable as
\eqn\pro{Z_+ = \{Q', \xi_+\} = e^{\phi^+} (d_+ - \gt_+ b) +
c\p\xi_+,}
$$
Z^{ab} = \{Q', \xi^{ab}\} = e^{\phi^{ab}} (d^{ab} -\half(\l^+)^{-1}\e^{abcde}\l_{cd} 
d_e) + c\p\xi^{ab},$$
$$
\widetilde Z^+ = \{Q', \widetilde\xi^+\} = e^{\widetilde\phi} (G^+ -\l^+ b)
+ c\p\widetilde\xi^+.$$

By inserting products of
these picture-raising operators, one finds
that the measure factors of \nat\ and \purem\ can be related as
\eqn\measure{Z ~
c \p c \p^2 c ~(\t)^{16}~
e^{-3\widetilde\phi -\phi^+ -\sum_{ab}\phi^{ab}} = 
(\l\g^m\t)(\l\g^n\t)(\l\g^p\t)(\t\g_{mnp}\t) + ...}
where
\eqn\zzz{Z = (\widetilde Z^+)^3 Z_+ \prod_{ab} Z^{ab} }
$$= b\p b\p^2 b ~\e^{\a_1 ... \a_{16}}
(\l\g^m)_{\a_1}(\l\g^n)_{\a_2}
(\l\g^p)_{\a_3}(\g_{mnp})_{\a_4\a_5} ~d_{\a_6} ... d_{\a_{16}}~e^{3\widetilde\phi +\phi^++
\sum_{ab}\phi^{ab}} + ...,$$
and $...$ involves terms with fewer than three $\l$'s (and more $c$'s)
and can be determined from the requirement of BRST invariance with respect to $Q'$.

\subsec{Scattering amplitudes}

For tree amplitudes which involve only the 
$GSO(+)$ vertex operators $V'$ defined in \vpr, the $N$-point
tree amplitude prescription is 
\eqn\treeplus{A = 
\prod_{r=4}^{N}\int dz_r
\langle\langle U_r(z_r) \prod_{s=1}^3 V'_s(y_s)\rangle\rangle
=
\prod_{r=4}^{N}\int dz_r
\langle\langle b(z_r) \prod_{s=1}^N V'_s(y_s)\rangle\rangle }
where $\int dz_r b(z_r)$ is the usual $b$ ghost insertion coming from
the Faddeev-Popov gauge-fixing of the worldsheet action.
Since there are no $\bt^+$'s in this correlation function, the only terms in
$V'$ of \vpr\ which contribute are $V' = c U + V$ and it is easy to verify
that \treeplus\ reproduces the original prescription of \treea.
It is interesting that, except for the different measure factor,
the prescription of \treeplus\ looks very
similar to the Lee-Siegel prescription of \ref\lstwo{K. Lee and
W. Siegel, {\it Simpler superstring scattering}, JHEP 0606 (2006) 046,
hep-th/0603218.} and it would be nice to find a proof that the two
prescriptions are equivalent.

But for tree amplitudes involving $GSO(-)$ vertex operators, one needs
to insert additional picture-changing operators to 
absorb the $(-{3\over 2},-\half,-\half)$ 
picture
of the $GSO(-)$ vertex operators of \tach. This procedure is precisely
analogous to RNS amplitudes involving Ramond states where
the number of picture-changing operators depends on the number of Ramond
vertex operators in the $-\half$ picture.

For example, for tree amplitudes involving $N$ $GSO(+)$ states $V'_{+}$
and
$2M$ $GSO(-)$ states $V'_{-}$, the tree amplitude prescription is
\eqn\treeminus{A = 
\prod_{r=4}^{N +2M}\int dz_r
\langle \langle b(z_r) Z^{M}(u)  \prod_{s=1}^N {V'}_{s+}\prod_{t=1}^{2M}
{V'}_{t-}
\rangle\rangle}
$$=
\prod_{r=4}^{N +2M}\int dz_r
\langle b(z_r) Z^{M-1}(u)  \prod_{s=1}^N {V'}_{s+}\prod_{t=1}^{2M}
{V'}_{t-}
\rangle$$
where the location of the picture-raising operators is arbitrary.
So for tree amplitudes involving two $GSO(-)$ states and an arbitrary number
of $GSO(+)$ states, one can use the natural measure factor of \nat\ without any
picture-changing insertions.

For genus $g$ amplitudes, the natural measure factor based on the
screening charges is
\eqn\natg{\langle b^{3g-3} (\t)^{16} (p)^{16 g}\exp [(g-1)(3\widetilde\phi + \phi^+ +\sum_{ab}
\phi^{ab})]\rangle =1.}
So one expects naively that the multiloop amplitude prescription
for $N$ $GSO(+)$ states and $2M$ $GSO(-)$ states is
\eqn\gplus{A = \prod_{j=1}^{3g-3}\int d\tau_j
\prod_{r=1}^{N +2M}\int dz_r
\langle\langle  b(z_r) b(\mu_j) 
Z^{M+g}(u)  
\prod_{s=1}^N {V'}_{s+}\prod_{t=1}^{2M}
{V'}_{t-}
\rangle\rangle}
where $b(\mu_j)$ is the $b$ ghost associated with the $j^{th}$
Teichmuller parameter $\tau_j$. When $M=0$, this prescription appears to be
closely related to the multiloop prescription given in \ref\multi{N. Berkovits,
{\it Multiloop amplitudes and vanishing theorems using the pure spinor 
formalism for the superstring}, JHEP 0409 (2004) 047, hep-th/0406055.}
for the
pure spinor formalism. However, a proof of equivalence of these
multiloop prescriptions will not be attempted here.

\newsec{Equivalence to Green-Schwarz Formalism}

In this section, the BRST operator $Q'$ of \nnq\ will be obtained
by gauge-fixing the Green-Schwarz superstring. But before discussing
the superstring, it will be useful to first discuss the Brink-Schwarz
superparticle. 

\subsec{Brink-Schwarz superparticle}

The $N=1$ $d=10$ Brink-Schwarz superparticle action, 
$S= \half \int d\tau ~e^{-1} \Pi^m \Pi_m$, can be written in first-order
form as \ref\brink{L. Brink and J.H. Schwarz, {\it Quantum superspace}, Phys. Lett.
100B (1981) 310.}\ref\siegelone{W. Siegel, {\it Hidden local supersymmetry in the
supersymmetric particle action}, Phys. Lett. 128B (1983) 397.}
\eqn\bsa{S = \int d\tau (P_m \pt x^m + p_\a \pt \t^\a -\half e P_m P^m 
+ f^\a d_\a)}
where $\Pi^m = \pt x^m + \half \t\G^m \pt\t$, $d_\a = p_\a -\half P^m (\G_m\t)_\a$,
and $f^\a$ is a fermionic Lagrange multiplier.

As is well-known, $d_\a=0$ contains 8 first-class constraints and
8 second-class constraints, and the first-class constraints are 
generated by 8 of the 16 components of the $\k$-symmetry generators
$P_m (\G^m d)^\a$. One can choose $G^A = \half(\G^+ \G^m d)^A P_m$
to describe these 8 first-class constraints where $A=1$ to 8
is an $SO(8)$ chiral spinor index, $\dot A=1$ to 8 is an $SO(8)$
antichiral spinor index, $J=1$ to 8 is an $SO(8)$ vector index,
and $\G^\pm \equiv \G^0 \pm \G^9$.
Note that $\{G^A, G^B\} = -\half\d^{AB} P^+ P^m P_m$.

Assuming that $P^+$ is nonzero, one can use $G^A$ to gauge-fix
$(\G^+ f)^A=0$ and can use the $P^2=0$ constraint to gauge-fix 
$e=0$. In this gauge, the BRST operator is
\eqn\qsp{Q =-\half c P^m P_m + \g_A G^A -\half P^+ \g_A \g_A b}
with the action
$$S = \int d\tau [
P_m \pt x^m + p_\a \pt \t^\a + b \pt c + \b^A \pt \g_A +
f^{\dot A} d_{\dot A}]$$
where 
$(\b^A,\g_A)$ are bosonic ghosts coming from the gauge-fixing of
$f^A$=0, and $f^{\dot A}d_{\dot A} = f^\a (\G^-\G^+ d)_\a$
describe the remaining second-class constraints.

To complete the BRST quantization, one needs to express the 8 second-class
constraints $d_{\dot A}=0$ in terms of 4 first-class constraints. This
is done by first splitting the eight components of $\g^A$ as
\eqn\splitg{\g_A = \d_A^+ \gt_+ + (P^+)^{-1} \l_A}
where $\l_A$ is a null $SO(8)$ spinor satisfying $\l_A \l_A =0$.
More explicitly, one decomposes the $SO(8)$ spinor $\g_A$ into $U(4)$
components as $\g_A \to (\g_+, \g_{jk}, \g_-)$ where $j=1$ to 4,
and defines 
\eqn\pured{\l^+\equiv\l_- = P^+ \g_-,\quad \l_{jk} = P^+ \g_{jk},}
$$
\l_+ = -{1\over 8} (\l^+)^{-1} \e^{jklm} \l_{jk}\l_{lm},\quad
\gt_+ = \g_+ - (P^+)^{-1} \l_+.$$
In terms of $\gt_+$ and $\l^A$, the BRST operator and action of \qsp\ are
\eqn\termb{Q =-\half c P^m P_m + \gt_+ G^+ + (P^+)^{-1} \l_A G^A - \gt_+ \l^+ b}
and 
\eqn\termc{S = \int d\tau [
P_m \pt x^m + p_\a \pt \t^\a + b \pt c + \bt^+ \pt \gt_+ +w_A\pt\l^A +
f^{\dot A} d_{\dot A}].}

One then defines the first-class constraints as
\eqn\fourf{H^J = \l\G^-\G^J d = \l^A \s^J_{A\Ad} d^\Ad}
where $\s^J_{A\Ad}$ are the $SO(8)$ Pauli matrices.
Note that $\l^A\l^A=0$ implies that
only 4 of the 8 components of $H^J$ are independent.
And since the components of $H^J$ anticommute with each other
and with the BRST operator of \termb, they can be used to replace
the 8 second-class constraints $d_\Ad =0$.

So one can replace \termc\ with the action
\eqn\termd{S = \int d\tau [
P_m \pt x^m + p_\a \pt \t^\a + b \pt c + \bt^+ \pt \gt_+ +w_A\pt\l^A +
h_J H^J]}
where only four components of the Lagrange multipliers $h^J$ are nonzero (e.g. choose
$h^5=h^6=h^7=h^8=0$). Note that the action of \termc\ is recovered if one uses the first-class
constraints of \fourf\ to gauge $d_\ad=0$, which produces no new propagating ghosts.\foot{I thank Yuri Aisaka for discussions on this point.} 
However, one can also use \fourf\ to gauge 
$h^J=0$, in which case the resulting BRST operator and action are
\eqn\termbb{
Q =-\half c P^m P_m + \gt_+ G^+ + (P^+)^{-1} \l_A G^A - \gt_+ \l^+ b + \g_J H^J,}
$$S = \int d\tau [
P_m \pt x^m + p_\a \pt \t^\a + b \pt c + \bt^+ \pt \gt_+ +w_A\pt\l^A + \b_J \pt\g^J]$$
where $\g_J$ are bosonic ghosts with only four nonzero components.
Finally, since $G^A = P^+ d^A + P^J \s_J^{A\Ad} d_\Ad,$
the BRST operator of \termbb\ is equal to the superparticle BRST operator
in the pure spinor formalism,
\eqn\puresb{
Q =-\half c P^m P_m + \gt_+ G^+ + \l^\a d_\a - \gt_+ \l^+ b}
where $\l^\a d_\a = \l^A d_A + \l^\Ad d_\Ad$ and
$\l_\Ad$ is defined as 
\eqn\ldeee{\l_\Ad \equiv (\g_J + P_J)\s^J_{A\Ad} \l^A}
which has four independent components.
Note that $\l^\Ad \l^\Ad = \l^A \l^A = \l^A \s^J_{A\Ad}\l^\Ad =0,$
so $\l^\a$ is a pure spinor with 11 independent components.

\subsec{Green-Schwarz superstring}

To extend these results from the Brink-Schwarz superparticle to the Green-Schwarz
superstring, first write the Green-Schwarz action in first-order form as \ref\gs
{M.B. Green and J.H. Schwarz, {\it Covariant description of superstrings}, Phys. Lett.
136B (1984) 367}
\siegeltwo
\eqn\GSa{S = \int d^2 z [ \half \p x^m \pb x_m + p_\a \pb \t^\a + 
\widehat p_\ah \p\widehat\t^\ah}
$$+ f^\a d_\a + \widehat f^\ah \widehat d_\ah - e (\half \p x^m \p x_m + p_\a \p\t^\a)
- \widehat e (\half \pb x^m \pb x_m + \widehat p_\ah \pb \widehat\t^\ah)]$$
where 
\eqn\defdd{d_\a = p_\a -\half\p x^m (\G_m\t)_\a -{1\over 8}
(\t\G^m\p\t)(\G_m\t)_\a, \quad \widehat d_\ah = \widehat p_\ah -\half\pb 
x^m (\G_m\widehat \t)_\ah
-{1\over 8}(\widehat\t\G^m\pb\widehat\t)(\G_m\widehat\t)_\ah,}
$f^\a$ and $\widehat f^\ah$ are fermionic Lagrange multipliers, $e$ and $\widehat e$
are the off-diagonal components of the worldsheet metric, and $(\a,\ah)$
are spinor indices of the opposite/same chirality for the Type IIA/IIB superstring.
In the following discussion, only the unhatted variables will be gauge-fixed, however, one
can gauge-fix the hatted variables in an identical manner.

As in the superparticle, $d_\a$ contains 8 first-class and 8 second-class constraints.
The first-class constraints are generated by 8 of the 16 components of
$\Pi_m (\G^m d)^\a$ where $\Pi^m = \p x^m +\half\t \G^m \p\t$ is the supersymmetric
momentum. Choosing
\eqn\defgggg{\widetilde G^A = \half(\G^+ \G^m d)^A \Pi_m}
to describe the 8 independent first-class constraints, one follows the same steps
as in the superparticle and gauge-fixes $(\G^+ f)^A = e =0$.

As shown in \siegeltwo, $\widetilde G^A$ satisfies the Poisson brackets
\eqn\pois{\{\widetilde G^A(\s_1),\widetilde  G^B(\s_2)\}}
$$ = \d(\s_1 -\s_2)
[\p\t^{(A} \widetilde G^{B)} + \d^{AB} \p\t^C\widetilde G^C 
+ \d^{AB} \Pi^+ (-\half\Pi^m\Pi_m - d_\a \p\t^\a) +\half \d^{AB} d_\Ad \p d_\Ad].$$
So the BRST operator and action after gauge-fixing are
\eqn\brgg{Q = \int dz [ c(-\half\Pi^m \Pi_m - d_\a \p\t^\a + 
\b\p\g + \p(\b\g) - b \p c) -\half \Pi^+ \g^A\g^A b + \g^A\widetilde G^A }
$$ - (\g_A\p\t^{A})(\g_B \b^{B})
-\half (\g_A\g^A) (\b_B\p\t^B) ],$$
\eqn\agg{S = \int d^2 z [\half \p x^m \pb x_m + p_\a \pb \t^\a + b\pb c + \b_A\pb \g^A
+ f^\Ad d_\Ad],}
where the last term in \pois\ can be ignored since it is quadratic
in the second-class constraints.

As in the superparticle, the 8 second-class constraints $d_\Ad=0$ can be converted
into 4 first-class constraints by writing $\g_A = \d_A^+ \gt_+ + (\Pi^+)^{-1}\l_A$
where $\l_A\l_A=0$, and defining the first-class constraints as
\eqn\defggg{H^J = \l^A \s^J_{A\Ad} d^\Ad}
which has only four independent components.
%
It is not difficult to verify that $H^J$ anticommutes with $H^K$ and satisfies
\eqn\qcom{\{Q, H^J\} =  \p (c H^J ) + \gt_+ H^K (\G^K\G^J\p\t)^+ + 
(\Pi^+)^{-1}(\l^A \p\t^A) H^J,} so $H^J$ describe first-class constraints
which can replace the 8 second-class constraints $d_\Ad$.
After gauge-fixing the Lagrange multiplier $h^J=0$ as in the superparticle,
the BRST operator of \agg\ becomes
\eqn\brrr{Q = 
\int dz [ c(-\half\Pi^m \Pi_m - d_\a \p\t^\a + w_A\p\l^A + \b_J\p\l^J
+\bt^+\p\gt_+ + \p(\bt^+\gt_+) - b \p c)}
$$ - \gt_+\l^+ b + \gt_+ \widetilde G^+ + 
\l_A \widetilde G^A 
+ \g_J H^J + ... ]$$
where $...$ involves ghost-ghost-antighost terms multiplied by components of $\p\t^\a$.
Finally, after defining
$\l_{\Ad} = (\g_J +\Pi_J) \s^J_{A\Ad} \l^A$ as in the superparticle,
one obtains the pure spinor BRST operator of \nnq\
\eqn\brstp{Q = \int dz (c \widetilde T + \gt_+ G^+ + \l^\a d_\a - \gt_+\l^+ b + b c \p c)}
where 
$\widetilde T = -\half \p x^m \p x_m - p_\a \p\t^\a + \widetilde T_{pure}$,
and $G^+ = \widetilde G^+ -{1\over 4} N_{mn} (\G^{mn}\p\t)^+ - {1\over 4}J \p\t^+
- {1\over 4}\p^2\t^+$ is defined as in 
\defgg. Although $G^+ - \widetilde G^+$ can be determined
by computing the ghost-ghost-antighost terms
in \brrr, $G^+ - \widetilde G^+$ can also be indirectly
determined by requiring the nilpotence of $Q$.

\newsec{Mapping RNS into the Pure Spinor Formalism}

In this section,
the RNS BRST operator will be mapped into the pure spinor BRST operator
by a field redefinition which maps the RNS variables into Green-Schwarz-Siegel
variables. For states in the Neveu-Schwarz $GSO(+)$ sector, the RNS and pure spinor
vertex operators in the zero picture will then be mapped into each other. However, 
since Ramond states in the RNS formalism and $GSO(-)$ states in the pure spinor
formalism do not have vertex operators in the zero picture, there is no obvious
way to map their vertex operators into each other.

\subsec{Twisting the RNS fields}

The first step in performing the map from the RNS BRST operator to the pure spinor
BRST operator is to twist the ten spin-half RNS fermions $\psi_m$ into five spin-zero
fermions $\widetilde\psi_m$ and five spin-one fermions $\widetilde{\bar\psi}_m$
as \baulieu
\eqn\twists{\psi^m ={1\over\g} \widetilde\psi_n {{(\l\G^m\G^n)^+}\over{2\l^+}} + 
\g\widetilde{\bar\psi}_n {{ (\l\G^n\G^m)^+}\over{2\l^+}} }
where $\g$ is the RNS bosonic ghost of $-\half$ conformal weight, 
$(\l\G^m\G^n)^+$ is the component of $(\l\G^m\G^n)^\a$ with ${5\over 2}$ $U(1)$
charge, and $\l^\a$ is a pure spinor which parameterizes the $SO(10)/U(5)$ 
different choices for twisting. Note that only 5 independent components of 
$\widetilde\psi_m$ and $\widetilde{\bar\psi}_m$ contribute to \twists, and \twists\
can be inverted to imply that
\eqn\twtw{\widetilde\psi_n (\l\G^m\G^n)^+ = \g \psi_n  (\l\G^m\G^n)^+,\quad
\widetilde{\bar\psi}_n (\l\G^n\G^m)^+ = {1\over \g}\psi_n  (\l\G^n\G^m)^+.}

Since the spin $3\over 2$ bosonic antighost $\b$ has non-trivial OPE's
with $\widetilde \psi_m$ and $\widetilde{\bar\psi}_m$, it
is convenient to define
new fields $\widetilde \b = \p\widetilde\xi e^{-\widetilde\phi}$
and $\widetilde \g = \widetilde \eta e^{\widetilde \phi}$ where
\eqn\newk{\widetilde \eta = e^{-\half \phi}\l^\a \Sigma_\a , \quad
\widetilde \xi = e^{\half \phi} (\l^+)^{-1} \bar\Sigma^+,}
$$
e^{\widetilde \phi}= \eta\p\eta e^{{5\over 2} \phi}
(\l^+)^{-1} \bar\Sigma^+ , \quad
e^{-\widetilde \phi}= \xi\p\xi 
e^{-{5\over 2}\phi} \l^\a \Sigma_\a, $$
and $\Sigma_\a$ and $\bar\Sigma^\a$ are anti-Weyl and Weyl spin fields
of ${5\over 8}$ conformal weight which are constructed in the usual
manner from the $\psi^m$ variables.
The definitions of \newk\ are uniquely determined by the requirements
that 
$[\widetilde\eta,\widetilde\xi, e^{\widetilde\phi}, e^{-\widetilde\phi}]$
have the same OPE's as 
$[\eta,\xi, e^{\phi}, e^{-\phi}]$ with each other, that
$[\widetilde\eta,\widetilde\xi, e^{\widetilde\phi}, e^{-\widetilde\phi}]$
have no poles with $\widetilde \psi_m$ and
$\widetilde {\bar\psi}_m$, and that $\widetilde\eta$ has
$+1$ conformal weight.  
Note that
\eqn\gam{
\widetilde\g =\widetilde\eta e^{\widetilde\phi} = \eta\p\eta e^{2\phi} = \g^2}
carries spin $-1$
and $\widetilde\b $ carries spin $2$. 
So the twisting of \twists\ and \newk\ shifts the central charge contribution
of the $(\b,\g)$ ghosts from 11 to 26, which cancels the shift from
$5$ to $-10$ in
the central charge contribution of the twisted $\widetilde\psi$'s.

If $\l^\a$ is treated as a worldsheet field, one needs to introduce
a fermionic superpartner for $\l^\a$ and add a topological term to the
RNS BRST operator so that these new fields do not contribute to the cohomology.
The fermionic superpartner to
$\l^\a$ will be called
$\widetilde\t^\a$ for reasons that will become clear, 
and will be defined to transform under BRST as
\eqn\fsp{Q\widetilde\t^\a = \l^\a, \quad Q\l^\a =0.  }
Furthermore, because of the pure spinor constraint $\l\G^m\l=0$, $\widetilde\t^\a$
will be required to satisfy the fermionic constraint
\eqn\fermc{\widetilde\t^\a\G^m_{\a\b} \l^\b =0.}
It is easy to verify that the constraint of \fermc\ eliminates five
components of $\widetilde\t^\a$, so that $\widetilde\t^\a$ and $\l^\a$
each have eleven independent
components.

To generate the BRST transformation of \fsp, one should add $\int dz
\l^\a \widetilde p_\a$ to the RNS
BRST operator so that 
\eqn\nonrns{Q =\int dz [c T_{RNS} - \g \p x^m\psi_m + b c\p c + \g^2 b + \l^\a \widetilde
p_\a]}
where $w_\a$ is the conjugate momenta to $\l^\a$
and $\widetilde p_\a$ is the conjugate momenta to
$\widetilde\t^\a$. Because of the constraints $\l\G^m\l = \l\G^m\widetilde\t=0$,
$w_\a$ and $\widetilde p_\a$ are defined up to the gauge transformations
\eqn\gauget{\d w_\a = \rho^m (\G_m \l)_\a + \Omega^m (\G_m\widetilde\t)_\a, \quad
\d \widetilde p_\a = \Omega^m (\G_m \l)_\a, }
where $\rho^m$ and $\Omega^m$ are arbitrary
gauge parameters. So five of the sixteen parameters of each of these
conjugate momenta can be gauged away.

To construct super-Poincar\'e covariant Green-Schwarz-Siegel variables
out of the RNS variables, one can now combine the
eleven components of $\widetilde\t^\a$ and $\widetilde p_\a$ with the five
spin-zero and spin-one components of $\widetilde\psi_m$ and $\widetilde{\bar\psi}_m$ to
define the unconstrained sixteen-component spinors
\eqn\pnew{\t^\a = \widetilde\t^\a + \widetilde\psi_m {{(\G^m)^{\a+}}\over{2\l^+}},\quad
p_\a = \widetilde p_\a + \widetilde{\bar\psi}_m (\G^m \l)_\a.}
Note that \pnew\
implies that $\psi^m$ can be expressed in terms of $p_\a$ and $\t^\a$ as
\eqn\psinew{ \psi^m = \g {{(\G^m p)^+}\over{2\l^+}} +
{1\over \g} (\l\G^m\t),}
and the OPE $\psi^m(y)\psi^n(z)\to (y-z)^{-1}\eta^{mn}$ implies that
$p_\a(y)\t^\b(z) \to (y-z)^{-1}\d_\a^\b.$

When expressed in terms of $p_\a$ and $\t^\a$,
\eqn\newhatg{\g \p x^m \psi_m = (\l\G_m \t)\p x^m
+\gt {{(\G_m p)^+}\over{2\l^+}} \p x^m.}
And $\l\G^m\l=0$ implies that $\l^\a p_\a =\l^\a \widetilde p_\a$.
So the BRST operator of \nonrns\ can be written as  
\eqn\qtwo{Q = \int dz [cT_{RNS} + b c \p c +
\l^\a ( p_\a - (\G^m\t)_\a \p x_m)
 + \widetilde\g
(b  - {{(\G^m p)^+}\over{2\l^+}} \p x_m)]}
where 
\eqn\trns{T_{RNS} = -\half \p x^m \p x_m + (\gt)^{-1} (\t\G^m\l)(\t\G_m\p\l)
-{{(\G^m p)^+}\over{2\l^+}} \p(\t\G_m \l)
+ \bt \p\gt + \p(\bt\gt) .}

Finally,
to put \trns\ into the standard form for a stress tensor and to covariantize
$(\G^m p)^+$ into $(\G^m d)^+$, 
one performs the similarity transformation
$Q\to e^U e^{S} Q e^{-S} e^{-U} $ where (up to possible errors in the coefficients)
\eqn\rnew{S = \int dz ~c [(\gt)^{-1}(\l\G^m\t)(\t\G_m \p\t) +
(\l^+)^{-1} (-{1\over 8}(\G^{mn}\p\t)^+(w\G_{mn}\l)
-{1\over 4} \p\t^+ (w\l))],}
$$U = \int dz ~(16\l^+)^{-1} (\G_{mn}\l)^+ \p x_p (\t\G^{mnp}\t).$$
After performing this similarity transformation,
\eqn\qsusy{Q = \int dz [ c\widetilde T + b 
c\p c + \l^\a d_\a + \gt 
(b -{{G^+}\over{\l^+}} ) ] = e^{R} [\int dz~(\l^\a d_\a + \gt b)] e^{-R}}
where $\widetilde T$, $G^+$, $d_\a$ and $R$ are defined as in section 2.  
So the RNS BRST operator has been mapped into the pure spinor BRST operator of \newq.

\subsec{Neveu-Schwarz $GSO(+)$ vertex operators}

In this subsection, the Neveu-Schwarz $GSO(+)$ vertex operators in the zero picture
in the RNS formalism will be mapped into the corresponding pure spinor vertex operators.
However, since Ramond vertex operators in the RNS formalism and $GSO(-)$ vertex operators
in the pure spinor formalism cannot be written in the zero picture, there is no obvious
way to relate the vertex operators for these states
in the two formalisms. Note that the map of
\newk\ acts in a simple manner on operators in the zero picture, i.e. operators which can be
expressed directly in terms of $\g$ and $\gt$. However, the map
acts in a complicated manner on operators in nonzero picture which contain
explicit $\phi$ or $\widetilde\phi$ dependence.

In the zero picture, unintegrated Neveu-Schwarz vertex operators in the RNS formalism
have the form
\eqn\nsv{V_{RNS} = \g W + c G_{-\half} W}
where $W$ is an $N=1$ superconformal primary of 
weight $\half$ constructed from $(x^m,\psi^m)$ and
$G_{-\half} W$ is the single pole of $\psi_m\p x^m$ with $W$.
After performing the field redefinition of \psinew, $V_{RNS}$ is expressed in terms of the
variables $[x^m, (\l\g^m\t),(\l^+)^{-1} (\g^m p)^+,\gt,c]$. And if the state is
$GSO(+)$, this operator contains integer powers of $\gt$.

To map $V_{RNS}$ to a pure spinor vertex operator, one needs to perform the similarity
transformation $V = e^{-R} e^U e^{S} V_{RNS} e^{-S} e^{-U} e^R$
where $R$, $S$ and $U$ are defined in \defR\ and \rnew. Since 
\eqn\sincee{e^{-R} e^{U} e^S (Q_{RNS} + \int dz~\l^\a p_\a)
e^{-S} e^{-U} e^R = \int dz (\l^\a d_\a + \gt b),}
$V$ is in the pure spinor cohomology.
But before claiming that $V$ is a pure spinor vertex operator, one needs to
ensure it is independent of inverse powers of $\l^+$ and $\gt$. One can show that 
any dependence on such inverse powers can be removed by adding a suitable
BRST-trivial operator, however, the form of this BRST-trivial operator may
be complicated to construct.

A more direct way to map
the Neveu-Schwarz $GSO(+)$ vertex operator of \nsv\
into the corresponding pure spinor vertex operator is to
write the $N=1$ superconformal primary $W$ of \nsv\
in the form
\eqn\formns{W = \psi^m f_m(x^n, M_{pq})}
where $M_{pq}= \psi_p\psi_q$ is the contribution
of $\psi^m$ to the RNS Lorentz current, and $f_m(x^n, M_{pq})$
is a function of $x^n$ and $M_{pq}$ and their worldsheet
derivatives. Since $GSO(+)$ superconformal primaries have an
odd number of $\psi$ fields, it is always possible
to write $W$ in the form of \formns\ for some choice
of $f_m(x^n,M_{pq})$.

The corresponding pure spinor vertex operator will then be defined as
\eqn\purev{V = (\l\g^m\t) f_m (x, M) + \sum_{n=1}^\infty V_{2n+1}}
where 
\eqn\mpre{M_{pq} = N_{pq} + \half (p\G_{pq}\t)}
and the terms in $V_{2n+1}$ contain $(2n+1)$ more $\t$'s than $p$'s.
Note that \open\ implies that $M_{pq}$ of \mpre\ has the same OPE's as $M_{pq}=\psi_p\psi_q$.
To determine the terms in $V_{2n+1}$, use $\{\int dz~\l^\a d_\a , V\}=0$
and 
$\{\int dz~\l^\a p_\a , (\l\g^m\t) f_m(x, M)\}=0$
to imply that
\eqn\vprimed{
\{\int dz ~\l^\a p_\a, ~ V_{2n+1}\} =}
$$
\{\half\int dz ~\l^\a \p x^m (\G_m\t)_\a, ~V_{2n-1}\} +
\{{1\over 8}\int dz ~\l^\a (\t\G^m\p\t)(\G_m\t)_\a, ~V_{2n-3}\} $$
where $V_1= (\l\g^m\t)f_m(x,M)$ and $V_{m}= 0$ for $m<0$.

Finding solutions to \vprimed\ for $V_{2n+1}$ would always be possible 
if $\int dz~\l^\a p_\a$ had trivial cohomology at $+2$ ghost number.
Although in fact there are non-trivial elements at $+2$ ghost number in the cohomology
of $\int dz~\l^\a p_\a$ (e.g. the state $(\l\G^m\t)(\l\G^n\t)(\t\G_{mnp}\t)$),
it seems reasonable to conjecture that when $W= \psi^m f_m(x,M)$ is an $N=1$
superconformal primary, these non-trivial elements are not an obstacle to finding
solutions for $V_{2n+1}$ which satisfy \vprimed.
So assuming this conjecture concerning $\int dz ~\l^\a p_\a$ cohomology at $+2$ ghost number,
there is a simple map from unintegrated Neveu-Schwarz $GSO(+)$ vertex operators in the RNS
formalism to unintegrated vertex operators in the pure spinor formalism.

One can similarly map integrated Neveu-Schwarz $GSO(+)$ vertex operators at zero
picture in the RNS
formalism into the corresponding pure spinor vertex operators. If
\eqn\integrr{\int dz~U_{RNS} = \int dz~ f(x,M)}
is the integrated vertex operator in the RNS formalism where $M_{mn}=\psi_m\psi_n$, then
\eqn\integp{\int dz~U = \int dz~ [f(x,M) + \sum_{n=1}^\infty U_{2n}]}
is the integrated vertex operator in the pure spinor formalism where
$M_{mn}=N_{mn} + \half(p\G_{mn}\t)$ and
$U_{2n}$ contains $2n$ more $\t$'s than $p$'s. In this case, finding solutions
to $U_{2n}$ is related to the cohomology of $\int dz~\l^\a p_\a$ at $+1$ ghost number.
When $\int dz~U_{RNS}$ is $N=1$ superconformally invariant, one expects that non-trivial
elements in this cohomology do not provide obstacles to solving for $U_{2n}$.

The maps of \purev\ and \integp\ can easily be verified for the massless
gluon vertex operator where $W=\psi^m a_m(x)$ and $U_{RNS}
= \p x^m a_m(x) + M^{mn} \p_m a_n(x).$ 
And since any massive Neveu-Schwarz $GSO(+)$
vertex operator can be obtained from the OPE's of gluon vertex
operators, this map is indirectly verified also for massive states. Furthermore,
since there are no terms in these vertex operators with more $p$'s than $\t$'s,
most of the terms $V_{2n+1}$ and $U_{2n}$ in the pure spinor vertex operators 
will not contribute. So using arguments similar to \ref\vall{N. Berkovits
and B.C. Vallilo, {\it Consistency of super-Poincar\'e covariant superstring tree amplitudes},
JHEP 0007 (2000) 015, hep-th/0004171.} one can verify 
that tree amplitudes involving Neveu-Schwarz $GSO(+)$ states coincide in the
two formalisms. However, since loop amplitudes involve intermediate states
in the Ramond $GSO(+)$ sector, it is not surprising that it is difficult to prove equivalence
of the RNS and pure spinor amplitude prescriptions for loop amplitudes.

\newsec{Inverse Map for $N=0 \to N=1$ Embedding}

After twisting the ten RNS spin-half fields into five spin-zero and five
spin-one fields, the RNS superstring was mapped in the previous section into
the pure spinor formalism. Since the pure spinor formalism can be interpreted
as an $N=2$ topological string \topo, which is a natural generalization of bosonic strings,
the map takes a critical $N=1$ string into a type of $N=0$ string.

As shown with Vafa \unique, any critical $N=0$ string can be embedded into a critical
$N=1$ string by twisting the $(b,c)$ ghosts from spin $(2,-1)$ to spin
$({3\over 2},-\half)$ and defining the $N=1$ superconformal generator as
$G=b+j_{BRST}$. In this section, it will be shown that if one starts with the
$N=1$ string corresponding to this $N=0\to N=1$ embedding of the bosonic string and
performs the map of the previous section, one recovers the original $N=0$ bosonic
string. 

So the map of the previous section from the RNS to the pure spinor formalism
can be interpreted as an inverse map for the $N=0\to N=1$ embedding of \unique.
This interpretation suggests there may be generalizations of the pure spinor formalism
which would arise by applying the inverse map to other types of critical $N=1$
superconformal field theories.

\subsec{Review of $N=0\to N=1$ embedding}

In this subsection, the map of \unique\ from a critical $N=0$ string to
a critical $N=1$ string will be reviewed. Suppose one starts with
a $c=26$ matter system with stress tensor $T_m$. Then the standard
quantization as a critical $N=0$ string
is to introduce $(b,c)$ ghosts of conformal weight $(2,-1)$
and define physical states using the $N=0$ BRST operator
\eqn\qnz{Q_{N=0} =  \int dz [ c T_m + b c \p c].}

However, the same matter system can also be quantized
as a critical $N=1$ string by adding a set of $(b_1, c_1)$ matter
fields of conformal weight $({3\over 2},-\half)$ so the combined
system has central charge $15$. One then defines a set of critical $N=1$
superconformal generators as
\eqn\superc{T_{N=1} = T_m - b_1 \p c_1 -\half \p(b_1 c_1) + \half \p^2
(c_1\p c_1),}
$$G_{N=1} = c_1 (T_m + \p c_1 b_1) + {5\over 2} \p^2 c_1 + b_1.$$
Note that $G_{N=1} = j_{BRST} + b_1$
where, up to a total derivative, $j_{BRST}$ is the BRST current of
\qnz\ with $(b,c)$ replaced by $(b_1,c_1)$.

One can now perform the standard $N=1$ quantization by
introducing fermionic $(b,c)$ ghosts of conformal weight $(2,-1)$
and bosonic $(\b,\g)$ ghosts of conformal weight $({3\over 2},-\half)$,
and defining physical states using the $N=1$ BRST operator
\eqn\qnf{Q_{N=1} =  \int dz [ c T_{N=1} +\g G_{N=1}
-\g^2 b +  b c \p c + c (\b\p\g +\half \p(\b\g))].}

Equivalence of the cohomologies of $Q_{N=0}$ of \qnz\ and $Q_{N=1}$ of \qnf\
was proven in \ref\hirosh{H. Ishikawa and M. Kato, {\it
Note on $N=0$ string as $N=1$ string}, Mod. Phys. Lett. A9 (1994)
725, hep-th/9311139.} by writing
\eqn\simo{Q_{N=1} = e^U [
\int dz (c T_m + b c \p c + \g b_1) ] e^{-U}}
where
$$U= \int dz c_1 (\half \g b - {3\over 2}\p c \b -  c\p\b
+\half \p c_1 c b -{1\over 4}\b\g\p c_1).$$
The topological term $\g b_1$ in \simo\ implies that the $N=1$
cohomology is independent of $(\b,\g)$ and $(b_1,c_1)$, so the
$Q_{N=1}$ cohomology coincides with the $Q_{N=0}$ cohomology.
Furthermore, it was shown in \unique\ that the $N=1$ amplitude prescription
coincides with the $N=0$ amplitude prescription
where the functional integral over the bosonic $(\b,\g)$ fields cancels
the functional integral over the fermionic $(b_1,c_1)$ fields.

\subsec{Inverse map for bosonic string}

In this subsection, it will be shown that if one starts with the $N=1$ string
coming from the $N=0\to N=1$ embedding of the bosonic string and performs similar
steps as in the map from the RNS to the pure spinor formalism, one 
ends up with the original $N=0$ description of the bosonic string.

The first step is to 
twist the $(b_1,c_1)$ matter fields from spin $({3\over 2},-\half)$ to
$(2,-1)$ by defining \baulieu
\eqn\twb{\widetilde b_1 = {1\over \g} b_1, \quad \widetilde c_1 = \g c_1}
as in the twisting of the $\psi^m$ matter fields in the RNS formalism.
Since $\b$ has non-trivial OPE's with $\widetilde b_1$ and $\widetilde c_1$, it is
convenient to define new fields $\widetilde \b = \p\widetilde\xi e^{-\widetilde\phi}$
and $\widetilde \g = \widetilde \eta e^{\widetilde \phi}$ where
\eqn\newh{\widetilde \eta = \eta e^{\half (\phi - i\s)}, \quad
\widetilde \xi = \xi e^{\half (-\phi+i\s)},}
$$
e^{\widetilde \phi}= \eta e^{\half (3\phi + i\s)}, \quad
e^{-\widetilde \phi}= \xi e^{\half (-3\phi - i\s)}, $$
and $c_1 = e^{i\s}$ and $b_1 = e^{-i\s}$.
The definitions of \newh\ are uniquely determined by the requirements
that
$[\widetilde\eta,\widetilde\xi, e^{\widetilde\phi}, e^{-\widetilde\phi}]$
have the same OPE's as
$[\eta,\xi, e^{\phi}, e^{-\phi}]$ with each other, that
$[\widetilde\eta,\widetilde\xi, e^{\widetilde\phi}, e^{-\widetilde\phi}]$
have no poles with $\widetilde c_1 = \eta e^{\phi+i\s}$ and
$\widetilde b_1 = \xi e^{-\phi-i\s}$, and that $\widetilde\eta$ has
$+1$ conformal weight.
One can easily verify from \newh\ that
\eqn\ghk{\widetilde\g = \eta\p\eta e^{2\phi} = \g^2 }
carries spin $-1$
and $\widetilde\b $ carries spin $2$.
So the twisting of \twb\ and \newh\ has shifted the spins of both $(b_1,c_1)$
and $(\b,\g)$ from $({3\over 2},-\half)$ to $(2,-1)$, and their contributions
to the central charge continue to cancel each other.

When written in terms of $(\widetilde b_1,\widetilde c_1)$
and $(\widetilde\b,\widetilde\g)$, $Q_{N=1}$ of \qnf\ becomes
\eqn\qthre{Q_{N=1} =
\int dz [ cT +
\widetilde c_1 (T_m + \p\widetilde c_1 \widetilde b_1)
+ \widetilde\g (b - \widetilde b_1) -\half \widetilde\g \p^2 ({{\widetilde c_1}\over
{\widetilde\g}}) + b c \p c]}
where
\eqn\tone{T=T_m +\half\p^2 ({{\widetilde c_1\p\widetilde c_1}\over{\widetilde\g}})
-\widetilde b_1 \p \widetilde c_1 -
\p(\widetilde b_1 \widetilde c_1) + \widetilde\b\p\widetilde\g +
\p(\widetilde\b\widetilde\g) .}
To put $T$ into the standard form for a stress tensor, one can perform
the similarity transformation $Q\to e^{S}Q e^{-S}$ where
$S= -\half (\widetilde\g)^{-1} \widetilde c_1 \p^2 \widetilde c_1$ which transforms $Q$ into
\eqn\transqq{ Q= \int dz [
c (T_m - \widetilde b_1\p\widetilde c_1 -\p (\widetilde b_1\widetilde c_1)
+ \widetilde \b\p\widetilde \g +\p (\widetilde \b\widetilde\g) )
+ \widetilde c_1 
(T_m - \widetilde b_1\p\widetilde c_1) + \widetilde\g (b-\widetilde b_1) + b c \p c]}
$$=e^{R}
\int dz [
\widetilde c_1 (T_m + \p\widetilde c_1 \widetilde b_1)
+ \widetilde\g b ] e^{-R} =
e^{R} (Q'_{N=0} + \int dz ~\widetilde\g b) e^{-R},$$
where $R= \int dz ~ c(\widetilde b_1 + \p c \widetilde \b)$
and $Q'_{N=0}$
is the
$N=0$ BRST operator of \qnz\ with
$(b,c)$ replaced by $(\widetilde b_1, \widetilde c_1)$.

Since $\widetilde\g b$
is a topological term, the twisted
$(\widetilde\b,\widetilde\g)$ ghosts will now cancel out
the contribution of the $(b, c)$ ghosts instead of the $(b_1, c_1)$
matter fields.
The remaining fields include the $c=26$ matter fields and the
$(\widetilde b_1,\widetilde c_1)$ matter fields of $(2,-1)$ conformal weight, which
are treated like $(b,c)$ ghosts in the standard $N=0$
description. So this inverse map
takes the $N=1$ description of the bosonic string into the
$N=0$ description.
     
\newsec{Comparison with Other Approaches}

In this paper, many mysterious features of the pure spinor formalism
were explained by adding a pair of non-minimal fields and performing
a similarity transformation which allows the pure spinor BRST operator
to be expressed in a conventional-looking form. Although this approach
is the first one that has succeeded in describing the $GSO(-)$ sector,
there have been
several previous approaches to ``explaining'' the pure spinor formalism
and it will be useful to compare this paper with the other approaches.

One approach has been to relax the pure spinor constraint on the ghost
variable $\l^\a$ and extend the BRST operator to include additional terms
which are required for nilpotence \ref\grassi
{P.A. Grassi, G. Policastro, M. Porrati and P. Van Nieuwenhuizen,
{\it Covariant quantization of superstrings without pure spinor constraints},
JHEP 0210 (2002) 054, hep-th/0112162\semi
P.A. Grassi, G. Policastro and P. Van Nieuwenhuizen,
{\it The quantum superstring as a WZNW model}, Nucl. Phys. B676 (2004)
43, hep-th/0307056.}
\ref\yuri{Y. Aisaka and Y. Kazama, {\it A new first-class algebra, homological
perturbation and extension of pure spinor formalism for superstring},
JHEP 0302 (2003) 017, hep-th/0212316\semi
Y. Aisaka and Y. Kazama,
{\it Relating Green-Schwarz and extended pure spinor formalisms
by similarity transformation}, JHEP 0404 (2004) 070,
hep-th/0404141.}\ref\chester{M. Chesterman,
{\it Ghost constraints and the covariant quantization of the superparticle
in ten dimensions}, JHEP 0402 (2004) 011, hep-th/0212261.}
\ref\siegellee{K. Lee and W. Siegel, {\it Conquest of the ghost
pyramid of the superstring}, JHEP 0508 (2005) 102,
hep-th/0506198.}. Although the conventional-looking BRST
operator in this paper also includes additional terms, the extended
BRST operators generically require an infinite number of additional terms
in order to be nilpotent. It might eventually be possible to relate these
extended approaches with the approach of this paper, however, it seems to 
be much easier to work with the conventional-looking BRST operator which has
a finite number of terms. Even though the conventional-looking BRST operator
is not manifestly Lorentz invariant, it is easy to show that the resulting
scattering amplitudes are Lorentz invariant.

A second approach has been to derive the pure spinor formalism from
a semi-light-cone gauge-fixed version of the Green-Schwarz formalism which
has double the usual number of $\t$ variables \ref\marchioro
{N. Berkovits and D. Marchioro, {\it  Relating the Green-Schwarz and 
pure spinor formalisms for the superstring}, JHEP 0501 (2005) 018,
hep-th/0412198.}\ref\origin{Y. Aisaka and Y. Kazama, {\it
Origin of pure spinor superstring}, JHEP 0505 (2005) 046,
hep-th/0502208.}. The resulting equivalence proof
with the GS formalism is certainly related to the proof in section 4 of this paper,
however, the equivalence proof in this paper is considerably simpler
and does not require the choice of semi-light-cone gauge.

A third approach has been to interpret the pure spinor formalism as
a topological string \topo\ and to compute scattering amplitudes by coupling
to worldsheet topological gravity 
\sken. Although this approach is 
probably not
useful for comparing with the RNS and GS formalisms, it might
eventually be useful for constructing generalizations of the pure
spinor formalism, perhaps by looking for other examples of the
$N=1\to N=0$ inverse map of section 6.

Finally, a fourth approach has been to relate the pure spinor formalism
with the $N=1\to N=2$ embedding of the RNS string \relatr\ref\tonin
{M. Matone, L. Mazzucato, I. Oda, D. Sorokin and M. Tonin,
{\it The superembedding origin of the Berkovits pure spinor
covariant quantization of superstrings}, Nucl. Phys. B639 (2002)
182, hep-th/0206104.}.
For compactification of the superstring on a Calabi-Yau manifold,
this $N=1\to N=2$ embedding is related by a field redefinition to the
hybrid formalism \ref\hybridfour{N. Berkovits,
{\it Covariant quantization of the Green-Schwarz superstring in a 
Calabi-Yau background}, Nucl. Phys. B431 (1994) 258,
hep-th/9404162.}. And in ten dimensions, this $N=1\to N=2$
embedding is related by a field redefinition to the GS ``twistor string''
\tonin\ref\tonintwo{M. Tonin, {\it Worldsheet supersymmetric
formulations of Green-Schwarz superstrings}, Phys. Lett. B266 (1991)
312.}\ref\mehet{N. Berkovits, {\it The heterotic Green-Schwarz
superstring on an $N=(2,0)$ worldsheet}, Nucl. Phys. B379 (1992) 96,
hep-th/9201004.}.
If this fourth approach were better understood, it might lead to a proof
of equivalence of the RNS and pure spinor multiloop amplitude prescriptions.
Furthermore, this approach might allow compactifications of the pure
spinor formalism to be related to the hybrid formalism.
However, there are some unresolved puzzles concerning this approach.

One puzzle is that the pure spinor formalism appears to be described
by a topological $N=2$ string which has $\hat c=3$, and
not by a critical $\hat c=2$ $N=2$ string which arises from the
$N=1\to N=2$ embedding. Note that naive compactification of the pure
spinor formalism produces a $\hat c=3$ $N=2$ theory which, unlike
the hybrid formalism, only describes the BPS sector of the compactified
superstring \topo. Also, the string field theory action for the pure spinor
formalism resembles a Chern-Simons action, as opposed to the 
Wess-Zumino-Witten-like action 
\ref\wzw
{N. Berkovits, {\it Super-Poincar\'e invariant superstring field theory},
Nucl. B450 (1995) 90, hep-th/9503099.}
which naturally arises from the
$N=1\to N=2$ embedding.

Nevertheless, as will be discussed in the following subsection,
there is a version of the $N=1\to N=2$ embedding which has many
similarities with the fields appearing in the $N=1\to N=0$ embedding
and which may eventually be useful for relating the pure spinor
and hybrid formalisms. The possibility of using this version of the
embedding to relate the pure spinor and hybrid formalisms has been
independently observed by Osvaldo Chand\'{\i}a \ref\ochan{O. Chand\'{\i}a,
private communication.}.

\subsec{$N=1\to N=2$ embedding}

The hybrid formalism for the superstring is constructed by first
embedding the RNS string into a $\hat c=2$ $N=2$ string,
and then finding a field redefinition which maps the RNS variables
into super-Poincar\'e covariant Green-Schwarz-Siegel variables \newdes.
The untwisted $\hat c=2$ $N=2$ generators are defined in terms of
the RNS fields as
\eqn\ntwo{T = T_{RNS} - \half \p J,\quad
G = j_{BRST},\quad \bar G = b,\quad J= cb + \eta\xi,}
where $j_{BRST}$ is the RNS BRST current and $(\xi,\eta)$ come
from fermionizing the $(\b,\g)$ ghosts as $\b=\p\xi e^{-\phi}$
and $\g=\eta e^\phi$.
The field redefinition to Green-Schwarz-Siegel variables is then
defined by 
\eqn\fdo{\t^\a = e^{\phi\over 2}\Sigma^\a,\quad p_\a =
e^{-{\phi\over 2}} \Sigma_\a}
where $\Sigma^\a$ is the RNS spin
field of conformal weight $5\over 8$. This field redefinition can
be used for the subset of $\t^\a$ variables which are
chosen in the $+\half$ picture.

However, one can also consider the field redefinition \baulieu
\eqn\fdt{\t^a \equiv \widetilde\psi^a = \g \psi^a,\quad p_a \equiv\widetilde{\bar\psi}_a
= (\g)^{-1} {\bar\psi}_a}
where $a=1$ to 5, $\psi^a = {1\over{\sqrt 2}}(\psi^a +i\psi^{a+5})$, 
and 
${\bar\psi}_a = {1\over{\sqrt 2}}(\psi^a -i\psi^{a+5})$. 
This field redefinition
is related to \twists\ by a fixed choice of $\l^\a$ in which the only nonzero
component is $\l^+$. In terms of $\t^a$ and $p_a$, the $\hat c=2$
$N=2$ generators of \ntwo\ are
\eqn\nthr{T =- \p x^a \p \bar x_a - p_a \p\t^a + \bt\p\gt +
\p (\bt\gt) - b\p c - \p(bc) - \half \p J,}
$$G = c (T +\half \p J) +b c \p c + \gt p_a \p x^a + \t^a \p \bar x_a +\gt b,$$
$$\bar G = b,$$
$$J = cb + 2 \bt\gt + \t^a p_a,$$
where $x^a ={1\over{\sqrt 2}}( x^a + i x^{a+5})$, $\bar x_a = 
{1\over{\sqrt 2}}(x^a -i x^{a+5})$, and
$\gt = (\g)^2$. 
As in the hybrid formalism, all variables in \nthr\ are automatically $GSO$-projected
so there is no need to sum over spin structures.

Finally, performing the similarity transformation $\phi \to e^{-R}\phi e^R$
on all worldsheet fields $\phi$ where
\eqn\rrdef{R = \oint dz ~(c p_a \p x^a + c \p c \bt),}
one can express the $N=2$ generators as\ref\slide{N. Berkovits,
Strings 2002 proceedings (Univ. of Cambridge),
http://www.damtp.cam.
ac.uk/strings02/avt/berkovits/img\_16.html .}
\eqn\nfour{T = -\p x^a \p \bar x_a - p_a \p\t^a + \bt\p\gt +
\p (\bt\gt) - b\p c - \p(bc) - \half \p J,}
$$G =  \t^a \p \bar x_a +\gt b,$$
$$\bar G = - p_a \p x^a + \bt\p c + \p (\bt c) +b,$$
$$J = \t^a p_a + cb + 2 \bt\gt.$$
Note that the $b$ ghost in $\bar G$ is not necessary for closure of the $N=2$
superconformal algebra, and
if one ignores the presence of the $b$ ghost in $\bar G$, \nfour\
are the standard $N=2$ generators for a set of 5 chiral and 
antichiral scalar superfields, $(x^a,\t^a)$ and $(\bar x_a, p_a)$,
and a set of spin $-1$ chiral and spin $2$ antichiral superfields,
$(c,\gt)$ and $(\bt,b)$. 

Since this version of the $N=1\to N=2$
embedding contains similar fields to the $N=1\to N=0$ map to the pure spinor formalism,
it may be useful for proving the equivalence of the hybrid
and pure spinor formalisms. Note that unlike the usual hybrid
formalisms defined using \fdo, the $N=2$ generators of \nfour\
do not involve chiral bosons. For example, the $N=2$ generators
in the $d=4$ hybrid formalism \hybridfour\
involve a chiral boson $\rho$. These chiral bosons have been
an obstacle to computing multiloop amplitudes using the hybrid
formalism and it is possible this new version of the $N=1\to N=2$
embedding will be useful for computing multiloop amplitudes which
can be compared with the multiloop prescription in the pure spinor
formalism. 

Furthermore, one can easily compactify this version of the embedding
on a Calabi-Yau $D$-fold
by replacing $\sum_{a=1}^5 \t^a \p\bar x_a$ with
$\sum_{a=1}^{5-D} \t^\a \p\bar x_a + G_C$ and
$\sum_{a=1}^5 p_a \p x^a$ with 
$\sum_{a=1}^{5-D} p_a \p x^a + \bar G_C$ where
$(G_C,\bar G_C)$ are the fermionic $N=2$ generators of
the Calabi-Yau $D$-fold.
Finally, the fact that $(x^a,\t^a)$ and $(\bar x_a,p_a)$
appear in the $N=2$ generators of \nfour\ in the same manner
as they appear in topological strings suggests there may be
a close connection between this version of the $N=1\to N=2$
embedding and topological strings.

\vskip 20pt
{\bf Acknowledgements:} I would like to thank 
Yuri Aisaka, Osvaldo Chand\'{\i}a, 
Nikita Nekrasov and Warren
Siegel for useful discussions, and 
CNPq grant 300256/94-9
and FAPESP grant 04/11426-0 for partial financial
support.

\listrefs
\end